\documentclass[12pt]{iopart}
\usepackage{epsfig}
\usepackage{indentfirst}
\usepackage{ulem}
\usepackage{color}

\begin{document}

\title[Semiclassical model for calculating ionization FDCS of H$_2$ molecule]
{Semiclassical model for calculating fully differential ionization cross sections
of the H$_2$ molecule}

\author{F J\' arai-Szab\' o\dag, K Nagy-P\'ora\dag\ and L Nagy\dag}

\address{\dag\  Faculty of Physics, Babe\c{s}-Bolyai University, str.
Kog\u{a}lniceanu 1, RO-400084 Cluj-Napoca, Romania}
\ead{jferenc@phys.ubbcluj.ro}
\begin{abstract}
Fully differential cross sections are calculated for the ionization of H$_2$ by fast charged projectiles
using a semiclassical model developed previously for the ionization of atoms.
The method is tested in case of 4 keV electron and 6 MeV proton projectiles.
The obtained results show good agreement with the available experimental data.
Interference effects due to the two-center character of the target are also observed and analyzed.
\end{abstract}

\submitto{\JPB}

\section{Introduction}
In the last few decades there was a considerable development in the theoretical description
and experimental measurement of differential cross sections for charged particle impact ionization
of atoms and molecules \cite{Erhardt1986,Ullrich1997,Roder1998,Dorner2000}.
Nowadays, the interest is focused on the in-detail
analysis of the electron ejection from atomic or molecular targets
\cite{Ullrich2003,Schulz2003,Schulz2004,Dimopoulou2004,Casagrande2008,Baran2008}.
These analyzes may be performed by measuring and calculating fully differential cross sections
which gives us the most complete
information about an ionization process. These quantities describe the entire energy and angular
distribution of the ionized electron, residual ion and projectile.

Previously, based on the semiclassical impact parameter method we have constructed a theoretical model
to calculate fully differential cross sections for single ionization of light atoms.
This model takes into account the projectile--target nucleus interaction classically.
The method
was tested in case of the single ionization of helium produced by C$^{6+}$ ion projectile with an
energy of $E_0 = 100$ MeV/u and good agreement with the experiments
was achieved in the scattering plane, while in the perpendicular plane a structure similar to that
observed experimentally was obtained \cite{Jarai2007,Jarai2009}.

More complex and interesting features appear when the target is a molecule. Due to the
multicenter character of the target, interference patterns occur in the ejected electron spectra.
This phenomenon was observed in the double differential cross section (DDCS) data for the
ionization of hydrogen molecule by fast ions by
Stolterfoht \etal \cite{Stolterfoht2001} and has been analyzed theoretically by several groups
\cite{Galassi2002,Laurent2002,Nagy2002,Stia2003,Sarkadi2003}.

Recently, fully differential cross sections were measured and calculated
for the ionization of H$_2$ by both fast and
slow electrons \cite{Casagrande2008,Stia2002,AlHagan2008}, and
interference effects were analyzed in these data series. 
However, while several attempts have been made in the description of the interference 
effects in the double differential cross sections, and these effects are relatively well 
understood \cite{Galassi2002,Nagy2002}, it was shown recently \cite{Laforge2009}, that an
accurate description of the role of the projectile--target nucleus interaction remains a major
challenge to theory.
There are even much more details to clarify for these effects, if they appear in
the total differential cross section.

Interference effects due to indistinguishable diffraction of the incoming projectile
from the two atomic centers were identified in the DDCS as a function of scattering angle for fixed
ejected electron energy \cite{Alexander2008}.
It was shown,
that these kinds of interference structures can be more pronounced, that those in the ejected
electron spectrum.

The goal of the present paper is to adopt our previous semiclassical method \cite{Jarai2007}
to calculate fully differential cross sections (FDCS) for the ionization of the hydrogen molecule,
and to evidence the interference effects due to the two-center character of the target.
In this model, the description of the molecule is similar to that from paper \cite{Nagy1992a},
where total cross sections are calculated and are compared to the experimental data.
Due to the limitations of a semiclassical model, we study the ionization induced by
fast charged projectiles, and analyze the interference structures only in the ejected
electron spectra.

As a test case, the single ionization
of the hydrogen molecule by 4 keV electron and 6 MeV proton projectiles is considered.
FDCS are calculated and compared in the scattering plane with the experimental data of
Cherid \etal \cite{Cherid1989} and Dimopoulou \etal \cite{Dimopoulou2004}.
Results in the perpendicular plane are also presented. In order to
evidence the interference effects due to the two-center character of the target, the fully differential
ionization cross sections for the H$_2$ molecule and H atom are compared.

\section{Semiclassical theory}
In the semiclassical approximation the projectile is treated
separately and it moves along a classical trajectory. This implies that only the electron system needs to be
described by a time-dependent Schr\"odinger equation, while the projectile follows the classical laws of
motion.
\subsection{General theory}
In order to study the ionization process of a small molecule induced by fast charged projectiles, first the
ionization amplitudes have to be calculated.

As described in \cite{Nagy2002}, the first-order transition amplitude for a projectile
with impact parameter $b$, velocity $v$ and charge $Z_p$, and a certain orientation of the
molecular axis $\mathbf{D}$ may be written as
\begin{equation}
a(b, \mathbf{k}, \mathbf{\hat D}) = i \frac{Z_p}{v} \int_{-\infty}^{+\infty}{dz\,e^{i\frac{\Delta E}{v} z}
\left\langle \psi_{\mathbf k}({\mathbf r}) \left| \frac{1}{\left| {\mathbf r} - {\mathbf R} \right|} \right|
\psi_i({\mathbf r}, {\mathbf D}) \right\rangle }\,,
\end{equation}
where $\mathbf r$ is the position vector of the active electron and $\mathbf R = b\,\mathbf{e_x} + z\,\mathbf{e_z}$ is
the position of the projectile along the trajectory with $z = vt$. Here the origin is considered in the center of the
molecule. The $\Delta E = E_i + \frac{k^2}{2}$ is the energy transfer to the active electron, where $E_i$ stands for
the ionization energy. The initial and final state of
the active electron is denoted by $\psi_i({\mathbf r}, {\mathbf D})$ and $\psi_{\mathbf k}({\mathbf r})$, with
$\mathbf k$ its momentum vector. In the calculations the Coulomb interaction
$\frac{1}{\left| {\mathbf r} - {\mathbf R} \right|}$ is expanded into a multipole series.

The transition probability for a given impact parameter and orientation of the molecular axis is the square
of the transition amplitude modulus
\begin{equation}
w(b, \mathbf k, \mathbf{\hat D}) = \left| a(b, \mathbf k, \mathbf{\hat D}) \right|^2\,.
\end{equation}
Because in experiments the molecular orientation is usually unknown, the transition probabilities are averaged
over molecular orientation
\begin{equation}
w(b, \mathbf k) = \frac{1}{4\pi} \int d\phi_D \int d\theta_D \sin{\theta_D}\,w(b, \mathbf k, \mathbf{\hat D})\,.
\end{equation}

Finally, the triple differential cross section (TDCS) for the electron ejected in the energy range $[E,E+dE]$
and into the solid angle $d\Omega_e$
and the projectile scattered into the solid angle $d\Omega_p$ can be expressed as
\begin{equation}
\frac{d^3\sigma}{d\Omega_e d\Omega_p dE} = \frac{b}{\sin{\theta_p}} \left|\frac{db}{d\theta_p}\right| w(b, \mathbf k)\,.
\label{FDCS3}
\end{equation}
Because some experimental data \cite{Dimopoulou2004} are differential relative to the perpendicular momentum transfer
$\mathbf{q_\perp}$, we calculate also a FDCS expressed in terms of $\mathbf{q_\perp}$ and not of the
scattering angle
\begin{equation}
\frac{d^5\sigma}{d\mathbf{q_\perp} d\mathbf{k}} = \frac{b}{q} \left| \frac{db}{dq} \right| w(b, \mathbf k)\,.
\label{FDCS5}
\end{equation}

An important part of the model is to assign impact parameter values to certain projectile scattering angles.
As described in detail in \cite{Jarai2009}, this may be achieved using the transverse momentum
balance \cite{Ullrich1997}, which states that the momentum transfer is the sum
of the transverse components of the electron's and residual ion's momenta. Further it is assumed, that the impact parameter
is related to the momentum transfer to the residual ion, and we take into account the projectile-electron interaction separately.
Thus, in this model, the projectile--target nucleus interaction is accounted for.
By this point of view two main collision types are possible: (a) binary collision where most of the
momentum transfer is taken by the electron and (b) recoil collision where most of the momentum transfer
is taken by the target nucleus. Accordingly, two different impact parameter values have to be used in these collision regimes.
The transition between these impact parameters is realized smoothly in the transition regions.

The projectile scattering is treated as a classical potential scattering problem in the field of the target system
with nuclear charge $Z_t$ \cite{Newton2002}.
The simplest way to include the effect of the electrons around the target nucleus is to consider the potential to be a product
of the Coulomb potential and the Bohr-type screening function \cite{Everhart1955}. Here it has to be mentioned, that using this
potential we assume that the projectile scattering is produced by a spherical potential. 
This implies that
in the description of the projectile scattering the two-center nature of the molecule is neglected. 
While in the analysis of the cross section as a function of the projectile scattering angle the correct description of the projectile--target nucleus interaction is important \cite{Laforge2009}, we assume that the present approximation does not influence
significantly the character of our results as a function of electron ejection angle.

\subsection{The particular case of hydrogen molecule target}
In the particular case of the hydrogen molecule the initial state is represented by a Heitler-London type molecular wavefunction.
Written only for the active electron
\begin{equation}
\psi_i({\mathbf r}, {\mathbf D}) = N \left(e^{-\alpha\left| \mathbf r - \mathbf D/2 \right|} +
e^{-\alpha\left| \mathbf r + \mathbf D/2 \right|}  \right)\,,
\end{equation}
where $\alpha$ is the effective charge, $\mathbf D$ is the vector associated to the internuclear distance and $N$ denotes
a normalization factor. In order to separate the dependence on the direction $\mathbf{\hat D}$ of the molecular axis,
this is expanded into a Legendre series \cite{Nagy1992a}. In this series each term is characterized by the quantum numbers $l_i, m_i$.
In the followings, the expansion coefficients depending only on the molecular distances will be denoted by $c_{l_i}(r,D)$.

The $\psi_{\mathbf k}({\mathbf r})$ continuum wavefunction is expanded into partial wave series, depending on angular
momenta $l_f, m_f$ and phaseshift $\sigma_{l_f}$.
The radial part $R_{l_f}(kr)$ is a wavefunction
of the continuum electron moving in the mean field of the residual H$_2^+$ molecular ion approximated by the potential
\begin{equation}
V(r) = \left\{ \begin{array}{ll}\frac{1}{r}\,, & r>\frac{D}{2} \\ \frac{2}{D}\,, & r\leq \frac{D}{2}\end{array} \right. \,.
\end{equation}
This wavefunction is calculated numerically.

By these considerations the ionization amplitude will be \cite{Nagy1992a}
\begin{eqnarray}
a(b, \mathbf k, \mathbf{\hat D}) = i \frac{(4\pi)^{\frac{3}{2}}}{v} \frac{Z_p N_i}{2 N_f} \times\nonumber\\
    		  \times\sum_{l_f l_c l_i} \frac{i^{-l_f} e^{-i\sigma_{l_f}}}{\sqrt{(2l_f+1)(2l_c+1)(2l_i+1)}} C^{l_f 0}_{l_c 0 l_i 0} \times \nonumber\\
    		  \times \sum_{l_f l_c l_i} C^{l_f m_f}_{l_c m_c l_i m_i}
    		  Y_{l_i m_i}(\hat D) Y^*_{l_f m_f}(\hat k) G^{m_c}_{l_f l_c l_i}(k,b,D)\,,
\end{eqnarray}
where $N_i$ and $N_f$ are the normalization factors of the initial and final state target wavefunctions, respectively.
The quantum numbers $l_c, m_c$ corresponds to the multipole expansion of the Coulomb interaction and
\begin{eqnarray}
G^{m_c}_{l_f l_c l_i}(k,b,D)&=&\int_{-\infty}^{+\infty} dz\,e^{i\frac{\Delta E}{v} z} Y_{l_c m_c}(\hat R)\times\nonumber\\
    		  &\times&\int_0^{\infty} dr\,r^2\,R_{l_f}(kr)\,\frac{r_<^{l_c}}{r_>^{l_c+1}}\,c_{l_i}(r,D)\,.
\end{eqnarray}
is the integral over the projectile trajectory.

Using this expression, the transition probability is calculated and averaged over molecular orientation.
The FDCS  for ionization of the hydrogen molecule will look as
\begin{eqnarray}
\sigma_{\rm FDCS} = K \left(\frac{N_i}{N_f}\right)^2\times\nonumber\\
\times \sum_{l_f l_c l_i l'_f l'_c} \frac{i^{l'_f - l_f} e^{i(\sigma_{l'_f} - \sigma_{l_f})}}
{(2l_i+1)\sqrt{(2l_f+1)(2l'_f+1)(2l_c+1)(2l'_c+1)}} \times\nonumber\\
\times C^{l_f 0}_{l_c 0 l_i 0} C^{l'_f 0}_{l'_c 0 l_i 0}\sum_{m_c m_i m'_c}
C^{l_f (m_c + m_i)}_{l_c m_c l_i m_i} C^{l'_f (m'_c + m_i)}_{l'_c m'_c l_i m_i}\times\\
\times Y_{l_f (m_c + m_i)}(\hat k) Y^{*}_{l'_f (m'_c + m_i)}(\hat k)
G^{m_c}_{l_f l_c l_i}(k,b,D) G^{*\,m'_c}_{l'_f l'_c l_i}(k,b,D)\,,\nonumber
\end{eqnarray}
where $K$ is a constant depending on projectile charge, velocity and scattering angle, target charge and ejected
electron momentum and it differs for the different kinds of FDCS mentioned earlier (see equations (\ref{FDCS3}) and (\ref{FDCS5})).
While our results will be represented scaled to unity at their highest value in the scattering plane,
in our following discussion the $K$ constant will play the role of a simple scaling constant.

\section{Results and discussion}

In order to test the validity of the results of the semiclassical model for the H$_2$ ionization, calculated FDCS
for fast electron and proton projectile impact are compared to the available experimental
data \cite{Dimopoulou2004,Cherid1989}.

\begin{figure}
	\begin{center}
		\epsfxsize=7.5cm \epsfbox{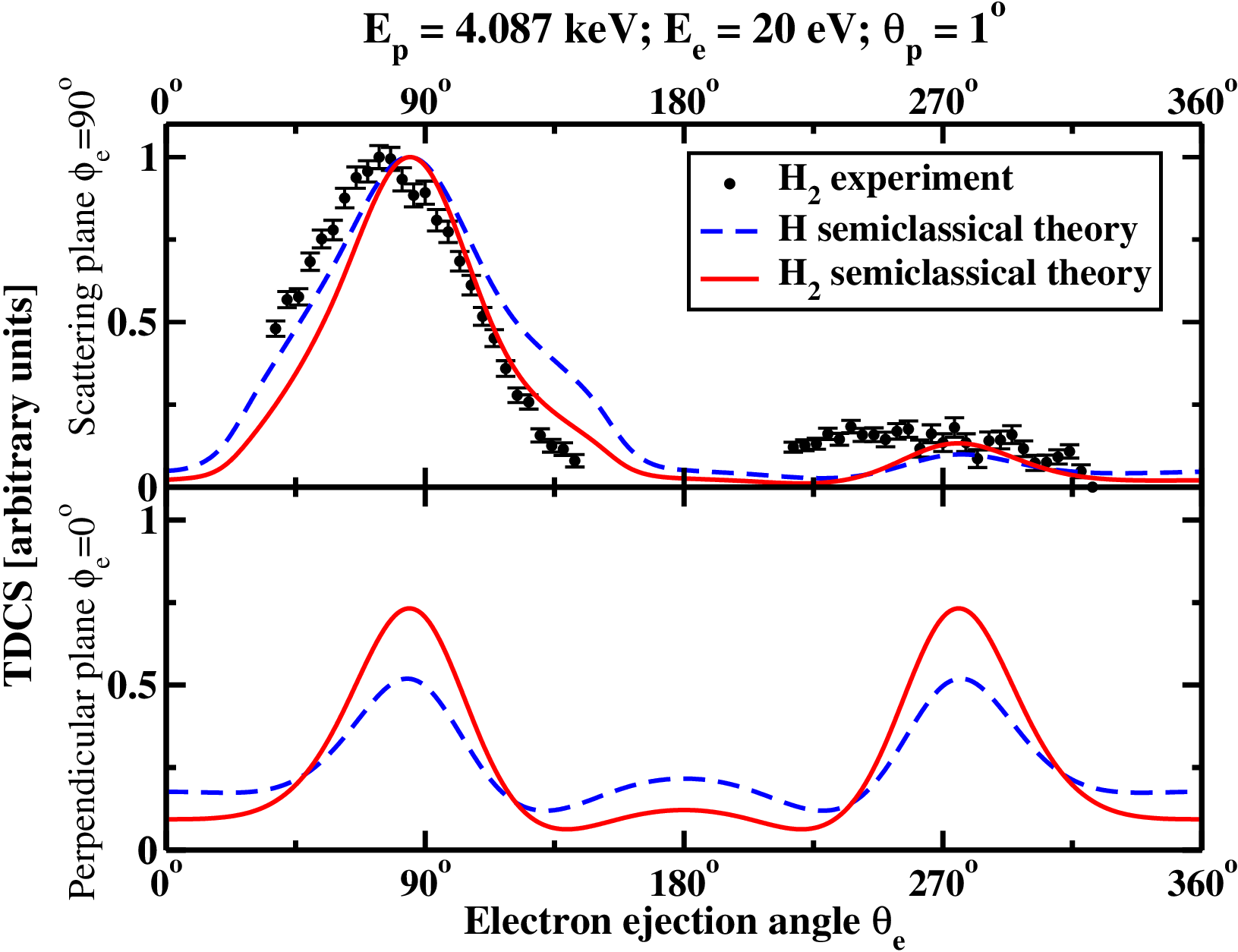}
	\end{center}
	\caption{\label{fig1}TDCS for ionization of H$_2$ molecule in scattering (top) and perpendicular (bottom) planes.
             Experimental results of Cherid et al \cite{Cherid1989} are represented by points while our
              semiclassical results for the H$_2$ molecule target are drawn with continuous lines. TDCS values for the
              same collision process with atomic H target are represented by dashed lines.}
\end{figure}

Fig. \ref{fig1} shows fully differential cross sections for the ionization of the hydrogen molecule by 4.087 keV electron
impact. The projectile scattering angle is $1^{\rm o}$ and the ejected electron energy is 20 eV.
The top panel of the figure shows TDCS in the scattering plane, while the bottom panel shows the same data in
the perpendicular plane.
Experimental and theoretical data are each scaled to unity at their highest value in the scattering plane.
Together with the results for the molecule (continuous line), TDCS are shown for the same collision process with atomic hydrogen target (dashed line).

Further comment needs the fact that the TDCS in the scattering plane for atomic hydrogen target show shoulder structures
at angles around 45$^\circ$ and 135$^\circ$. Such structures cannot occur for pure first-order ionization from an isotropic 1s state.
As presented above, in order to take into account the projectile--target nucleus interaction, a certain impact parameter is
assigned for every kinematic condition. Two different
impact parameter values are used in the two collision regimes. The transition between these impact parameters is realized
smoothly in the $\theta_e = 0^\circ ... 50^\circ$ and $\theta_e = 130^\circ ... 180^\circ$ transition regions. The shoulders appear due to this
impact parameter adaptation mechanism.

In the scattering plane the theory reproduces well in shape the binary and the recoil peak structure. However,
the semiclassical first-order results show a shift of the binary peak with some $10^\circ$
relative to the experimental data, which have the angular resolution of less than 1$^\circ$ \cite{Cherid1989}.
Such peak shifts indicates that the
first order approximation may not be sufficiently accurate.

Experimental data in the perpendicular plane for this particular process was not found. As in the case of our previous
studies of the He ionization \cite{Jarai2009}, the semiclassical theoretical results
show a double-lobe structure similar to that reported in the literature for slower collision processes \cite{AlHagan2008}.

In order to evidence the interference effects
due to the two-center nature of the target, TDCS for the H$_2$ molecule and H atom are compared in Fig. \ref{fig1}.
The only difference between these calculations is that the one-center wavefunction describing the H atom is replaced 
by a two-center wavefunction in case of the molecular target.
At the first look the TDCS distributions for H$_2$ molecule and H atom are very similar. However, the recoil
peak for the H$_2$ target is larger than that for the H atom. The interference effects are analyzed later by the means of
the interference factor. We note here, that we obtain the interference structures by comparing the
theoretical results for H$_2$ and H. Because the agreement between our calculations and the experimental data is not
perfect, we cannot say, that have identified the interference effects in the experimental data.

We have identified the interference effects in the electron ejection spectrum, due to the coherent ejection from the two centers of the molecule. This effect is obtained by describing the molecular electron by a two-center wavefunction. We assume, that neglecting the two-center character of the target in the description of the projectile scattering does not influence the obtained interference patterns.

The interference structures in the ejected electron spectrum are analyzed through
the interference factor defined by the ratio of the cross section obtained for the molecule and of two independent hydrogen
atoms \cite{Casagrande2008,Stia2003}
\begin{equation}
    I=\frac{\sigma_{\rm H_2}}{2 \sigma_{\rm H}}.
\end{equation}
This interference
effect may be detected in the TDCS distribution from non oriented molecules.
The interference factor may be expressed by an analytical approximate formula \cite{Casagrande2008}
\begin{equation}
    I = 1 + \frac{\sin(Dq^{'})}{Dq^{'}}\,,\label{analyticalI}
\end{equation}
where $\mathbf{q^{'}} = \mathbf{q} - \mathbf{k}$ is the momentum imparted to the recoil ion.

\begin{figure}
	\begin{center}
		\epsfxsize=7.5cm \epsfbox{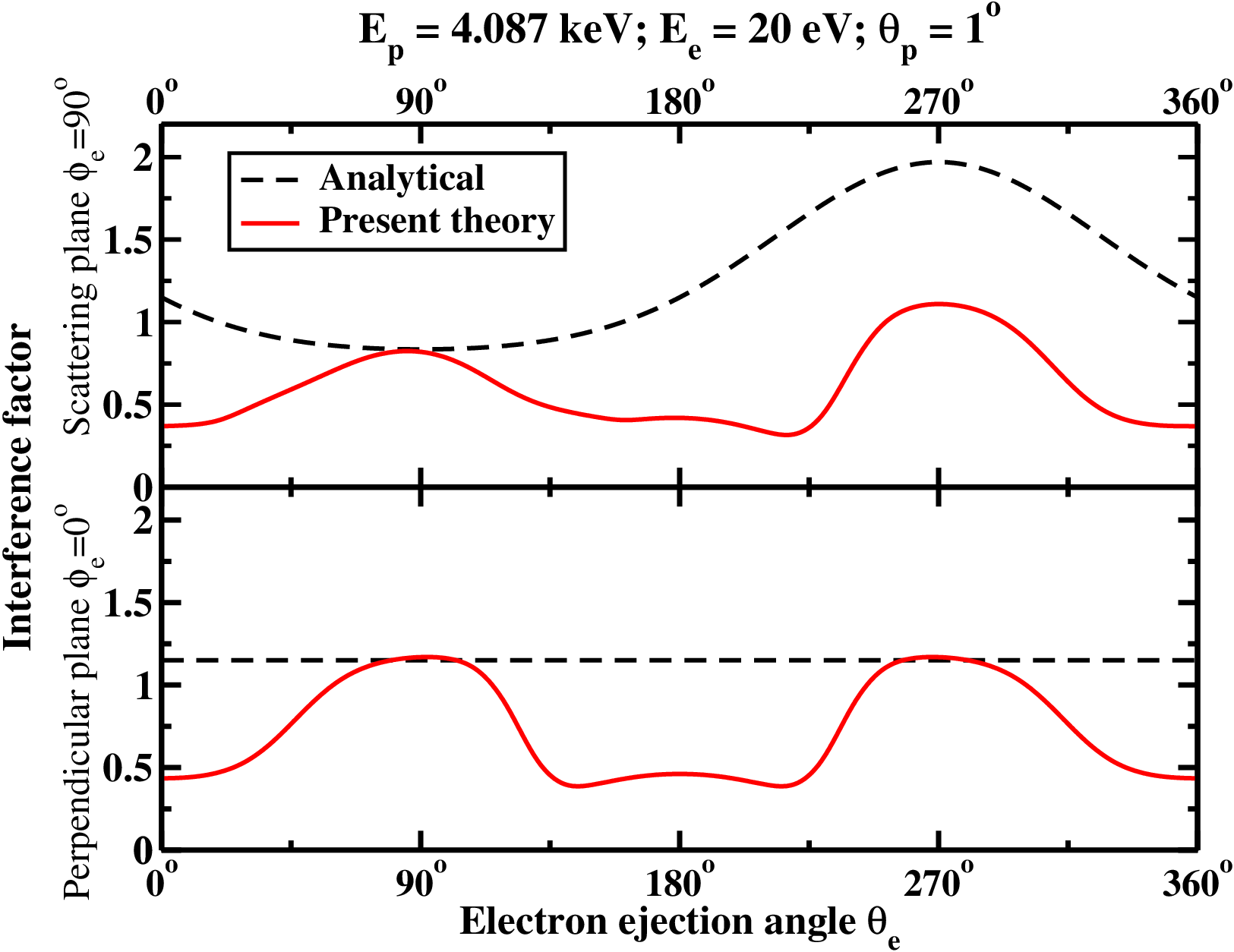}
	\end{center}
	\caption{\label{fig2}Interference factor for the process shown in Fig. \ref{fig1}.
            The continuous line represents results obtained by the present theory,
            while the dashed line shows the shape of the analytical form (\ref{analyticalI}).}
\end{figure}

In Fig. \ref{fig2} the interference factor obtained from our calculations (continuous line) is shown for the ionization
process discussed earlier and it is compared to the analytical form (\ref{analyticalI}) (dashed line). The top panel of the
figure shows the interference factor in the scattering plane. As expected, it has oscillatory behavior
with a strong maximum in the vicinity of the recoil peak. The semiclassical theory predicts another smaller maximum
in the binary peak region, too, suggesting a more complicated behavior than that given by the
analytical formula.
In the perpendicular plane due to the kinematic conditions the $\mathbf{q^{'}}$ momentum is constant:
(1) the transverse component of the recoil momentum $\mathbf{q^{'}}$ parallel to $\mathbf{q}$ is constant because
the electron momentum is zero in that direction and $\mathbf{q}$ is fixed; (2) the transverse component of
$\mathbf{q^{'}}$ perpendicular to $\mathbf{q}$ is constant because $\mathbf{q}$ is zero in that direction and
the electron energy is fixed. Accordingly, the analytical formula gives a constant interference factor.
In contrast to this result, the semiclassical theory predicts some oscillations symmetric relative to
the $180^{\rm o}$ direction (bottom panel of the Figure).

\begin{figure}
	\begin{center}
		\epsfxsize=7.5cm \epsfbox{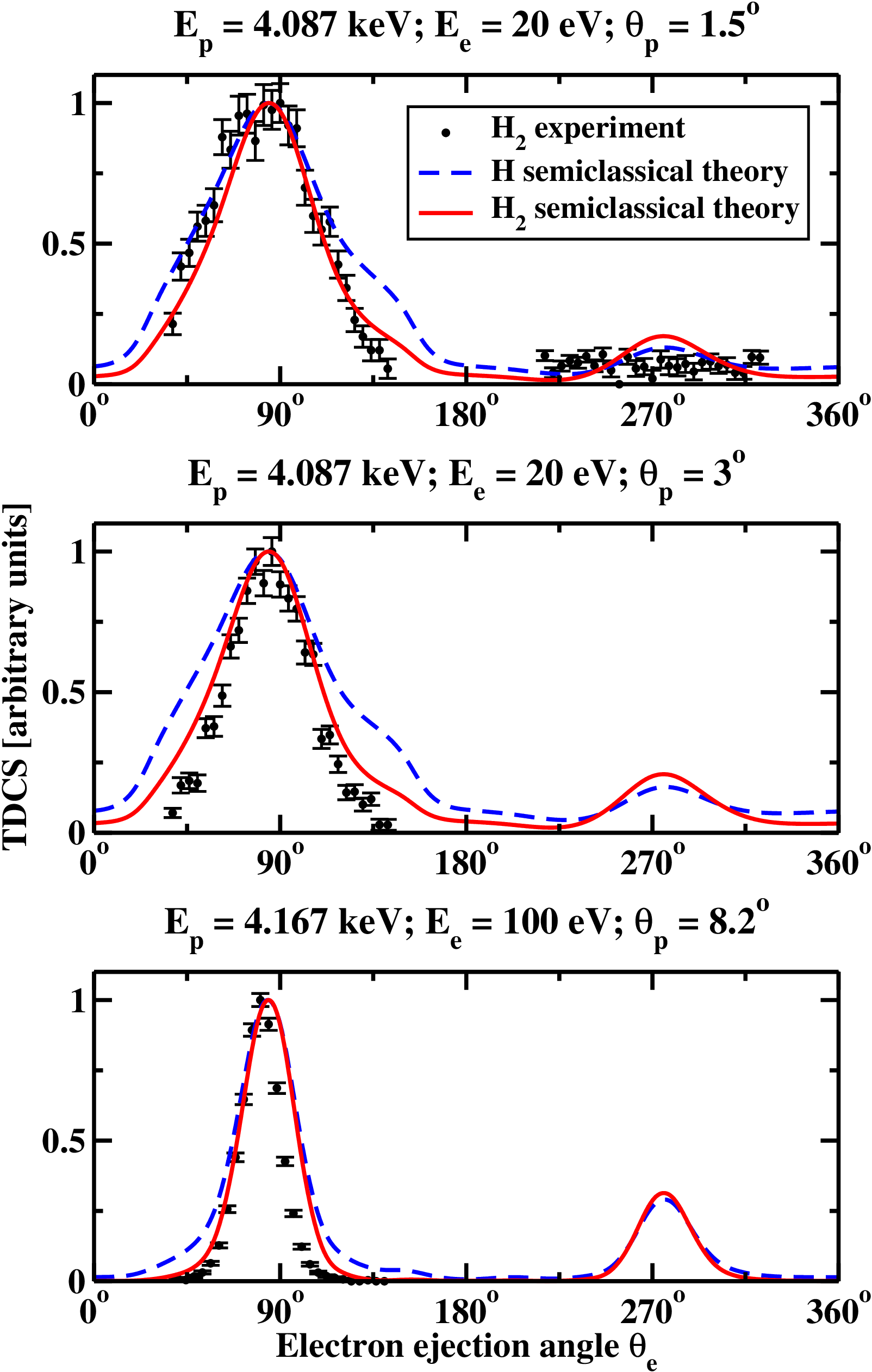}
	\end{center}
	\caption{\label{fig3}TDCS for ionization of H$_2$ molecule in scattering plane
	for different projectile energies, scattering angles and electron ejection energies.
   The legend of this figure is the same as of Fig. \ref{fig1}.}
\end{figure}

TDCS distributions for other projectile energies and scattering angles are analyzed in Fig. \ref{fig3}.
The top and middle panels present scattering plane TDCS results for 4.087 keV electron projectile with
$1.5^\circ$ and $3^\circ$ scattering angles.
The ionized electron is ejected with 20 eV energy. The results are in good agreement with experiments both in the
shape of the distribution and in the position of binary and recoil peaks.

The bottom panel of Fig. \ref{fig3} shows results for
4.167 keV projectile energy and a larger $8.2^\circ$ scattering angle. The electron is ejected with a higher energy, of
100 eV. In this case the experiments show a narrow binary peak, which is well reproduced by our calculations.

\begin{figure}
	\begin{center}
		\epsfxsize=7.5cm \epsfbox{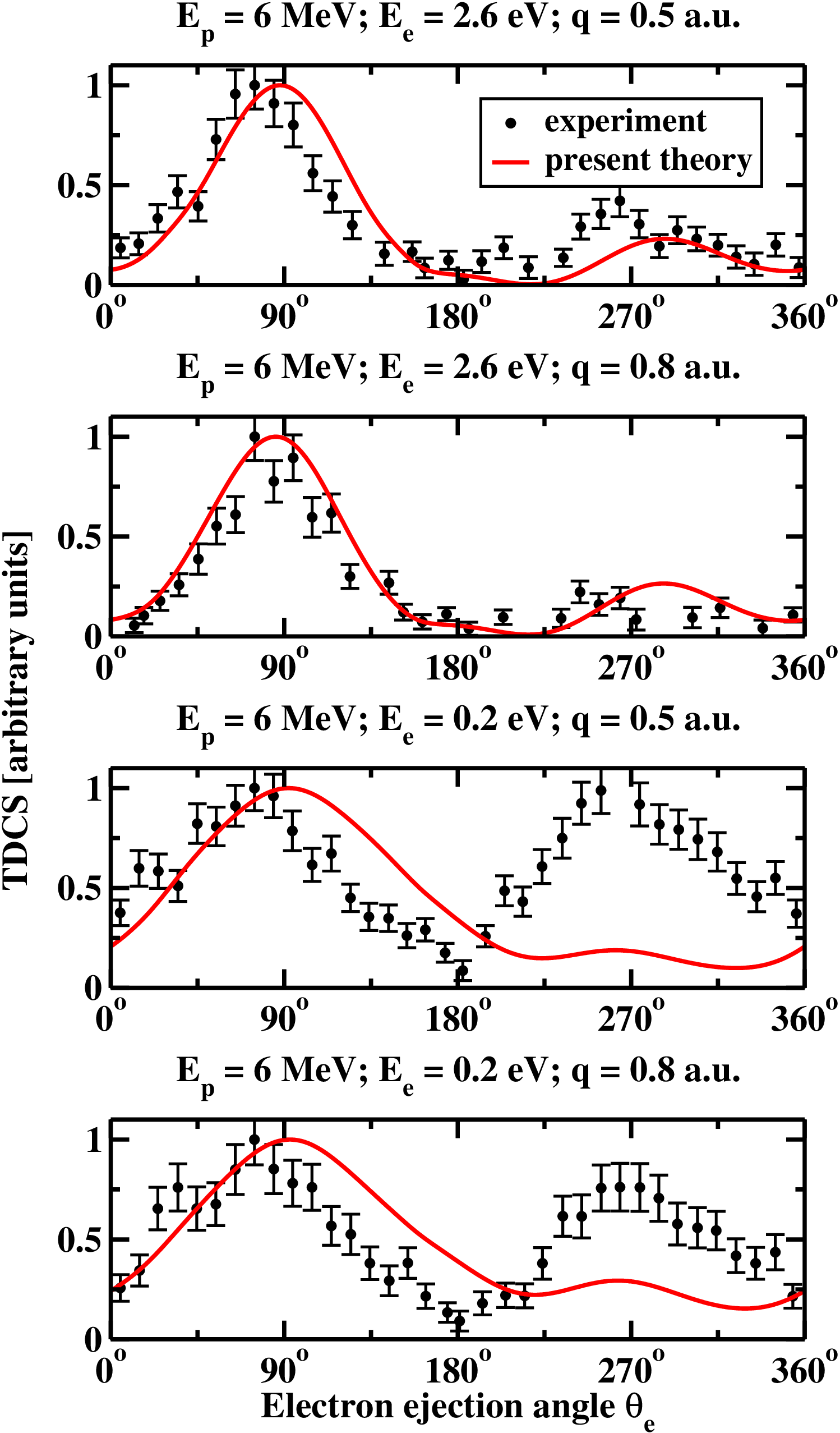}
	\end{center}
	\caption{\label{fig4}TDCS in scattering plane for the ionization of H$_2$ molecule by 6MeV proton projectile
	for different momentum transfers and electron ejection energies. As in previous figures
    the continuous line represents the results of the semiclassical theory, while dots are representing the
    experimental data of Dimopoulou et al \cite{Dimopoulou2004}.}
\end{figure}

Fig. \ref{fig4} shows FDCS results of the semiclassical model for ionization of hydrogen molecule
by 6 MeV proton projectile impact. In this case the experimental data of
Dimopoulou \etal \cite{Dimopoulou2004} is differential
in ejected electron momentum vector and the perpendicular momentum transfer vector. The comparison
is made for different electron ejection energies and momentum transfers.

On one hand, for larger electron energies, of $E_e = 2.6$ eV (top panels of the figure) the agreement
between the experimental data and the present, semiclassical theory is reasonably good.
Beside of this, it has to be noted that our first-order results show a shift
of the binary peak with some $10^\circ$, but because of the experimental
errors in the cross section this shift cannot be determined precisely.
The cross section ratios for the binary and recoil peaks are also well reproduced.

On the other hand, for lower electron energies, of $E_e = 0.2$ eV (bottom panels of the figure)
there is a discrepancy between the experimental and theoretical FDCS distributions. In experiments
the cross section ratio of the binary and recoil peaks is almost 1. This feature is not reproduced by
the semiclassical model. The explanation is the same as in the case of CDW-EIS model \cite{Dimopoulou2004}:
the FDCS distribution in the sub-eV region is influenced by the presence of the vibrational
autoionizing channel (not included into the semiclassical model) which leads to the ejection of
very low-energy electrons. The angular distribution of the autoionized electrons is
essentially a dipolar one with respect to the momentum transfer axis.

\section{Conclusions}

In conclusion, the theoretical model based on the first order, semiclassical,
impact parameter approximation used to calculate fully differential cross sections for single ionization
of light atoms has been adopted to calculate fully differential cross sections for the ionization of diatomic molecules.
The method has been tested in case of the ionization of H$_2$ by 4 keV electron and 6 MeV proton projectiles.
Except for some special cases, the obtained results in the scattering plane show good agreement with the experiments
\cite{Dimopoulou2004,Cherid1989},
and are in agreement with other theories \cite{Dimopoulou2004,Casagrande2008,Stia2002}. The double-lobe structure
reported in the perpendicular plane is also reproduced.
However, for low electron ejection energies the semiclassical model fails to reproduce
the experimentally observed more symmetrical electron emission patterns in the scattering plane.
In order to analyze the effect of the two scattering centers of the molecule on the cross section,
we have compared the TDCS for the H$_2$ with the TDCS obtained for independent atoms. Higher cross sections
were obtained for molecules in the region of the recoil peak, and also the width of the binary peak was changed.
These differences may be interpreted as interference effects.

The aim of this work was to test that the semiclassical model, in its simple form, is able to treat
complex colliding systems and describe interference patterns in the fully differential cross sections, too.

\section*{Acknowledgment}
The present work has been supported by the Romanian National Plan for Research (PN II),
contract No. ID 539.

\end{document}